\documentclass[referee]{aa} 
\usepackage{graphicx}
\usepackage{txfonts}
\usepackage{natbib}
\bibpunct{(}{)}{;}{a}{}{,}
\begin{document}

   \title{Carbon and oxygen isotopic ratios in Arcturus and Aldebaran:}
   \subtitle{Constraining the parameters for non convective mixing on the RGB}

   \author{C. Abia\inst{1}
          \and
          S. Palmerini\inst{1}
          \and
          M. Busso\inst{2}
          \and
          S. Cristallo\inst{3}
          }

   \institute{Dpto. F\'\i sica Te\'orica y del Cosmos, Universidad de Granada, E-18071 Granada (Spain)\\
              \email{cabia@ugr.es}
         \and
             Dipartimento di Fisica, Universit\'a di Perugia, and INFN, Sezione di Perugia, Italy\\
             \email{sara.palmerini@fisica.unipg.it;maurizio.busso@fisica.unipg.it}
         \and
             Osservatorio Astronomico di Collurania, INAF, Teramo 64100, Italy\\
             \email{cristallo@oa-teramo.inaf.it}
             }

\date{}

\abstract
{\bf We  re-analysed  the   carbon and oxygen  isotopic  ratios  in
the atmospheres of the two bright K giants Arcturus
($\alpha$ Boo) and   Aldebaran ($\alpha$ Tau).}
{\bf These  stars  are in the evolutionary stage following
the  first dredge-up (FDU). Previous determinations (dating more
than 20 years ago) of their $^{16}$O$/$$^{18}$O ratios showed a
rough agreement with FDU expectations; however, the estimated $^{16}$O$/$$^{17}$O and $^{12}$C$/$$^{13}$C
ratios were lower than in the canonical predictions for
red giants. Today these anomalies are interpreted as signs of the
occurrence of non-convective mixing episodes. We therefore
re-investigated this issue in order to verify whether the observed
data can be reproduced in this hypothesis and if the rather well
determined properties of the two stars can help us in fixing the
uncertain parameters characterizing non-convective mixing and in
constraining its physical nature.}
{\bf We used high-resolution infrared spectra from the
literature   to    derive  the    $^{12}$C$/$$^{13}$C   and
$^{16}$O$/$$^{17}$O$/$$^{18}$O ratios  from CO molecular lines
near 5  $\mu$m, using the LTE spectral  synthesis  method.  We made
use  of the  recently published ACE-FTS atlas of the  infrared
solar spectrum for constructing an updated atomic and
molecular  line lists in this spectral range. We also
reconsidered the determination of the stellar parameters to  build  the  proper  atmospheric
and evolutionary models.}
{\bf We found that both the C and the O isotopic ratios for the two
stars  considered actually disagree with pure FDU predictions.
This reinforces the idea that non-convective
transport episodes occurred in them. By reproducing the
observed elemental and isotopic abundances with the
help of parametric models for the coupled occurrence of
nucleosynthesis and mass circulation, we derived constraints on
the  properties of non convective mixing, providing information on
the so far elusive physics of such phenomena. We find that very slow mixing,
like that associated to diffusive processes, is incapable of explaining the
observed data, which require a rather fast transport. Circulation mechanisms
with speeds intermediate between those typical of diffusive and of convective
mixing should be at play.  We however conclude with a word of caution on the
conclusions possible at this stage, as the parameters for the mass transport
are rather sensitive to the stellar mass and initial composition.
At least for $\alpha$ Boo, reducing the uncertainty still remaining on
such data would be highly desirable.}
{}


   \maketitle
%

\section{Introduction}

Red giant branch (RGB) stars undergo evolutionary stages that
start at the so-called First Dredge-Up (FDU), a convective mixing
process carrying to the surface nuclei from internal layers,
previously affected by CN cycling. The FDU occurs as the He-core
contraction after the Main Sequence (MS)  is accompanied by a
downward envelope {\bf extension}. It is now well established  that this
leads to a  decrease in the  $^{12}$C/$^{13}$C ratio with respect
to the MS value  ($\sim 89$ in the  solar  case), down to values
in the range $15-30$ \citep[depending  on the initial  mass and
metallicity of  the star, see] []{wei00}.  In
addition,  the carbon abundance  drops in the envelope, while that
of nitrogen increases. If the stellar mass does not exceed $\sim
2$ M$_\odot$, the $^{16}$O abundance remains unaltered, while that
of $^{18}$O is mildly reduced. The  isotopic ratios
$^{16}$O/$^{17}$O/$^{18}$O expected by the models then {\bf lie} on a
characteristic line; their values depend on the initial stellar
mass and have moved strongly in recent years as a consequence of
changes in basic reaction rates \citep[see for example][especially their
Figure 3]{pal11y}.  The present situation for these ratios as a
function of the stellar mass, updated with the last version of the
FRANEC evolutionary code \citep{cris09,cris11} and with the last
recommendations available for the relevant reaction rates
\citep{ade11} is summarized in Table \ref{tab1}.

\begin{table}
\label{tab1}

\centering{
\caption{Oxygen isotopic ratios after FDU at solar metallicity}

\begin{tabular}{lcc} \hline\hline Mass (M$_{\odot}$) &
$^{16}$O/$^{17}$O & $^{16}$O/$^{18}$O\\ \hline 1&2571&526\\
1.2&2045&575\\ 1.25&1784&587\\ 1.3&1480&597\\ 1.4&1095&613\\ \hline
\end{tabular} }

\end{table}

This standard picture is challenged by a large amount of abundance
determinations \citep[see for example][]{bro89,gra00,char04,gru04} in
field and globular  cluster low-mass giant stars, showing very low
$^{12}$C/$^{13}$C ratios,  sometimes almost  reaching down to the
equilibrium value of the CN cycle  ($\sim 3.5$). Anomalies in the
C and O isotopes were found also in pre-solar C-rich and O-rich
grains of stellar origin, preserved in meteorites
\citep[for example][]{ama01,nit08}. In particular, Al$_2$O$_3$ grains
reveal remarkable $^{18}$O destruction \citep{nit97}.  Some
families of this cosmic dust display also isotopic shifts in
heavier elements including  Mg and Al  and others have anomalies
reaching up to neutron-capture elements beyond iron
\citep[for example][]{nico1,nico2}.

Observationally, evidence of anomalies is often found in low mass red giants
($\leq 2.3$ M$_\odot$) for phases subsequent to  the so-called {\it
Bump} of the Luminosity Function (BLF), when the advancing H-burning shell
erases the chemical discontinuity left behind by the first dredge up
\citep{Charbonnel2000}.  This homogenization facilitates the
occurrence of transport phenomena; hence, it became common to
attribute the chemical anomalies to the occurrence of episodes of
matter circulation in "conveyor belts"
\citep{Wasserburg1995,nol03,pal11y} or in diffusive processes
\citep{denis98,egg2006}. These phenomena (going under the names of
{\it deep mixing}, {\it extra-mixing}, or {\it cool bottom processes})
would link the envelope to regions where proton captures take place,
thus accounting for the observation that the photospheric material has
undergone extensive processing.

Among the proposed physical causes for mixing mechanisms one can
mention rotation itself through shear effects
\citep{zah92,wei00,char04} and meridional circulation
\citep{talon}; gravity waves \citep{den03}; magnetic buoyancy
\citep{bus07,den09}; and molecular weight inversions leading to heavier materials
falling down in a lighter environment \citep{egg2006}. This last process
was identified by \citet{ch07} as the known thermohaline double-diffusion
also occurring in the oceans and previously studied by \citet{ulrich} in astrophysical
environments.

Although  all  these  physical phenomena may have a role  in the complex dynamics linking red giant envelopes to their underlying radiative layers, it is not clear today in
which evolutionary stage each of them works more efficiently
\citep{Uttenthaler2007}, which is the range of stellar
masses affected and, therefore, which of them is more suited to
explain the abundance changes in RGB stars. Even the requirement that
the mixing episodes occur after the BLF has been recently questioned
\citep{drake}.

In general, the mentioned elusive processes are not treated in
canonical stellar models; in some cases they are intrinsically
linked to the stellar rotation or to the development of dynamical
instabilities, thus requiring at least {\bf two dimensional} hydro-codes
to be properly modeled. However, the use of 2D schemes for
general stellar evolution is (at best) in its infancy.

One has also to notice that clarifying the physics that is behind
the chemical peculiarities is made difficult by the fact that the
interpretation of observations is usually hampered by
uncertainties in fundamental parameters of the chosen stars
(stellar mass, luminosity, etc...). The sources we consider in
this paper, Arcturus ($\alpha$ Boo) and Aldebaran ($\alpha$ Tau),
are K-type RGB stars of nearly solar mass; they are very bright,
are situated at a limited distance from the Sun and are well
observed. Therefore, they may be less affected than others from
this last difficulty, and can be considered as good references for
studying stellar evolution and spectroscopic abundances in
first-ascent red giants. Indeed, the determination  of their
observed parameters (luminosity, radius and effective temperature)
is rather reliable and their large brightness facilitates the task
of getting high-resolution, high signal-to-noise-ratio spectra. This is so
for both optical and infrared wavelengths, thus allowing for an
accurate abundance analysis.

It has been known for decades that the  $^{12}$C/$^{13}$C ratios
of $\alpha$ Boo and $\alpha$ Tau share the problems discussed
above for common red giants, being lower than {\bf predicted} by the
first dredge-up (hereafter FDU).
\citet{hin76} and \citet{tom84} early derived ratios of 7 and 12
for these stars, respectively. These first estimates were then
confirmed by subsequent works \citep[see for example][]{smi90, pet93}.
Thus, the anomaly in their C isotopic ratios is a quite robust
result\footnote{The derivation of this ratio, mainly from optical
bands (CN lines in  the region $\sim 8000$ {\AA}) and from the
infrared domain (CO lines at $\sim 2.3~\mu$m),  is  rather
insensitive  to  the  stellar parameters adopted. Indeed, these
parameters  affect almost  equally the $^{12,13}$CN and/or
$^{12,13}$CO lines.}.

On the other hand, a previous determination of  the oxygen isotopic
ratios in these stars \citep{har84}, using CO lines at 2 and 5 $\mu$m,
indicated  $^{16}$O/$^{17}$O$\sim 1100$, $^{16}$O/$^{18}$O$\sim 550$
for Arcturus, and $^{16}$O/$^{17}$O$\sim 660$, $^{16}$O/$^{18}$O$\sim
475$ for Aldebaran.  The uncertainty was about $40-50\%$\footnote{We
report here only the O ratios derived by the above authors from the 5
$\mu$m region, as in the 2 $\mu$m region the available CO lines for K
giants are usually weak and blended.}. Actually, the mentioned authors
noticed the difficulty of explaining simultaneously the
$^{12}$C/$^{13}$C and $^{16}$O/$^{17}$O ratios in these stars within
the framework of the canonical FDU models. They proposed several
possible solutions: among others, strong mass loss prior to the FDU,
slow mixing during the main sequence and/or a reduction in the rate
for $^{18}$O($p,\alpha)^{15}$N.  No satisfactory solution was however
found, so that the problem remained open since then.

Recent theoretical calculations at solar metallicity
\citep{pal11y}, including revisions of critical nuclear rates
\footnote{Among which the new measurement of
$^{18}$O($p,\alpha)^{15}$N cross section provided by \citet{laco10}}, for masses close to 1.2 $M_{\odot}$, found O isotopic ratios
shown in Table 1\footnote{These predictions do not
change significantly with metallicity in the range
-0.5$\leq$[Fe/H]$\leq 0.0$.  (In the present work, we adopt the
standard notation [X/H]$=$ log(X/H)$_\star$-log(X/H)$_\odot$ where
(X/H) is the abundance of the element X by number in the scale log
(H)$\equiv 12$.)}.  A quick inspection to Table 1 reveals that, considering the observational uncertainty in
the Harris \& Lambert data for oxygen, the $^{16}$O/$^{18}$O
ratios in both stars  can be considered to be in rough agreement
with the new theoretical predictions of stars with $\sim 1.2$ M$_\odot$, but clearly this is not the case with the $^{16}$O/$^{17}$O ratio.  The purpose of this work
is to try solving this problem, possibly also deriving further
hints on the extra-mixing parameters that physical models must
reproduce.

In Sect. 2 we summarize  the input data for our analysis,  namely
the spectra we used, the stellar parameters and  the chemical
analysis tools we adopted.  Section  3  is  then devoted  to  the
description of the non-convective  models assumed for explaining
the newly determined C and O  isotopic ratios of our two program
stars. In Sect. 4 we then comment on the values found for the
extra-mixing parameters and we derive on this basis some general
conclusions.


\section{Input data}
\label{secdata}
\subsection{ Observed  data and  line lists} \label{sec21}
For  Arcturus we  used
the electronic  version of  the  {\it Infrared Atlas  Spectrum}�
by \citet{hin95}.  We analyzed the $\sim 5~\mu$m-region spectra
rationed to the telluric spectrum.  Accurate wavelength
positioning and identification  of the main species contributing
in this region were recently performed in the  solar infrared
spectrum by \citet{has10}. In particular, in the range  1800-2200
cm$^{-1}$  there are  many weak and unblended $^{12}$C$^{17}$O
and  $^{12}$C$^{18}$O lines very sensitive  to changes  of the O
isotopic ratios.  In the  case of Aldebaran, we  used a  spectrum
in a similar but shorter spectral region  obtained on  February 6,
1980 at the KPNO  4 m coud\'e telescope using a Fourier transform
spectrometer. This spectrum was kindly  provided by  K. Hinkle. It
has  a spectral  resolution of 0.016 cm$^{-1}$,  slightly lower
than that  of Arcturus  (0.01 cm$^{-1}$).  The spectrum  of
Aldebaran was  cleaned by telluric absorptions using the  telluric
spectrum  of  Arcturus (after  some spectral  resolution
degradation)  with the  IRAF task  {\it telluric}.   At  the wave
number position  of  the  strongest telluric absorptions  the
removal  was, however, unsatisfactory, thus these  spectral
regions were excluded in the  analysis. A difficulty in this
procedure is that  many peaks  are found  above the level of unity
in the rationed spectra. We detect these highest peaks (excluding
those close  to the regions  with the strongest  telluric
absorptions) and fit a  smooth curve passing through them
following the similar method used by  \citet{tsu09}  in  Arcturus
to  place  the continuum  level in  the rationed spectra. It is
uncertain whether the continuum adopted  in this  way  is a  true
continuum; however, different  fits to  these  peaks resulted  in
very small  differences ($\leq 1\%$) in the continuum level.

We made use  of an improved molecular line lists  in the $\sim 5~\mu$m
region. Our  list includes the molecules  CO, C$_2$, CN,  OH, SiO,
MgH, SiS, H$_2$O, being CO the main contributing  molecule in the
spectral region. CO lines come  from \citet{goo94}; C$_2$ lines are an
update  of \citet{que71} (private communication); CN lines  from B.
Plez (private communication), as an update after  the new energy
levels  calculated  by \citet{ram210} and \citet{ram10}; H$_2$O lines
are from \citet{bar06}; SiO ones from \citet{lan93}; MgH ones from
\citet{sko03}; OH ones from \citet{gol98}, SiS ones  from
\citet{cam09} and OH ones from the HITRAN database \citep{rot09}.  The
atomic  lines are taken from the VALD v-0.4.4 database \citep{kup00}.
A few line positions  and intensities (mainly CO lines)  were
corrected by  comparing a theoretical spectrum  of the Sun with the
infrared solar spectrum \citep{has10}. We used a MARCS atmosphere
model for the  Sun \citep{gus08}  with the solar abundances set from
\citet{asp09}. For the C and O isotopic ratios in the Sun we also
adopted the values  suggested by these authors. The fit to the  solar
spectrum  was  excellent, in  particular between  2100-2200 cm$^{-1}$
where  most of the $^{13}$C$^{16}$O, $^{12}$C$^{17,18}$O lines used to
derive the C and O ratios were selected.

\subsection{Atmosphere parameters} \label{sec22}
\begin{table}
\caption{Stellar parameters used to derive CNO abundances and
isotopic compositions in $\alpha$ Boo and $\alpha$ Tau.}
\label{phys} \centering
\begin{tabular}{lccc}
\hline\hline
 & $\alpha$ Boo & $\alpha$ Tau & Ref.\\
\hline
Name & Arcturus& Aldebaran & \\
HR & 5340  &  1457& \\
MK type &K1.5 III &  K5 III & \\
$T_{\rm eff}$[K] & 4290$\pm$ 50 & 3981$\pm$75 & a\\
$L$ [$L_{\odot}$] & 196$\pm$21 & 440$\pm$20& b\\
log\,($g$ [cms$^{-2}$])& 1.50$\pm$0.10 & 1.20$\pm$0.30 & a\\
$R$ [$R_{\odot}$] & 25.4$\pm$0.2  &    45.2$\pm 0.7$& a\\
$[$Fe/H$]$ & $-$0.50$\pm$0.07 &  $-$0.13$\pm$0.13 & a\\
$\xi_{micro}$[kms$^{-1}$]& 1.7$\pm 0.1$ &   1.94$\pm0.20$& a\\
$M$ [$M_{\odot}$] & 1.08$\pm$0.06 & 1.3$\pm$0.3 & b\\
\hline
\end{tabular}
\tablefoot{The following references are for Arcturus and
Aldebaran, respectively: \tablefoottext{a}{\citet{ryd09};
\citet{ram09} and \citet{alv10}} \tablefoottext{b}{\citet{ram11};
\citet[and references therein]{leb12}.} }
\end{table}

$\alpha$ Tau and $\alpha$ Boo have been extensively studied with
high resolution, high S/N spectra since 1980. Their atmospheric
parameters have been estimated by several authors using a variety
of techniques. For Arcturus, the simple mean and standard
deviation for the stellar parameters compiled in the PASTEL
database \citep{sou10} is T$_{eff}=4324 \pm 90$ K, log
g$=1.71\pm0.29$, and [Fe/H]$=-0.56\pm 0.1$, while for Aldebaran is
T$_{eff}=3850\pm 40$ K, log g$=1.2\pm 0.4$, and [Fe/H]$=-0.16\pm
0.1$. Despite the considerable number of studies on these stars,
the published parameters still do distribute randomly around the
mean values, due to the impact of systematic errors which vary
among different studies\footnote{A recent discussion on the
current techniques to derive atmospheric parameters and the
associated errors can be found in \citet{leb12}.}.  We decided,
therefore, to adopt the most recent determinations of the
atmospheric parameters for both stars. They are based on
high-quality visual and infrared spectra and the use of MARCS model
atmospheres (the same grid that we use here). For Arcturus we
adopted those derived in \citet{ryd09} while for Aldebaran those
from  \citet{ram09}, and \citet{alv10} (see Table \ref{phys}). The
adopted values, in any case, do not differ significantly from the
average values given in the PASTEL database. In the references
quoted above the C, N, O, abundances were derived, as well as
those for other species having an important role in the opacity of
the model atmospheres (Si, Ca, Mg, S, Ti etc, see the original
works for details). However, the atomic features present in the 5
$\mu$m region in both stars are very weak or severely blended;
therefore we adopt the elemental abundances given by the quoted
authors except for the CNO elements. We note that variations up to
$\pm 0.25$ dex in the metallicity of the stars has no impact in
the derived C and O isotopic ratios.

A spherical MARCS model atmosphere \citep{gus08} was interpolated
for each star from the grid of models for the parameters
(T$_{eff}$, log g, [Fe/H], $\xi_{micro}$) in Table \ref{phys}. We
assumed for both stars $1$ M$_\odot$ and for Arcturus we adopted
an $\alpha-$enhanced model ([$\alpha$/Fe]$=+0.4$) as suited to its
metallicity \citep{pet93}. We note that Arcturus has a
chromosphere \citep[for example][]{ayr75}, so that the continuum flux in
enhanced at wavelengths shorter than about 2000 ${\AA}$. This flux
excess might be explained using a binary model \citep{ver05}, but
in this study we considered Arcturus as a single star, as the
impact of the possible secondary companion is only important in
the ultraviolet, a region that we did not use and that therefore
does not affect our analysis.

\subsection{Analysis of the C and O ratios} \label{sec23}

\begin{table*}

\caption{CNO abundances and isotopic ratios} \label{tabobs}
\centering
\begin{tabular}{c c c c c c c c}     
\hline\hline
Star& log $\epsilon$(C) & log $\epsilon$(N) & log $\epsilon$(O) &  $^{12}$C/$^{13}$C&  $^{16}$O/$^{17}$O&  $^{16}$O/$^{18}$O & $^{17}$O/$^{18}$O\\
\hline                        
$\alpha$ Boo& $8.06\pm 0.09$ (20) & $7.67\pm 0.13$ & $8.76\pm 0.17$ & $ 9 \pm 2$ (24)& 3030$\pm 530$ (7)& 1660$\pm 400$ (18)& 0.55$\pm 0.12$\\
$\alpha$ Tau& $8.25\pm 0.12$ (15)  & $8.05\pm 0.11$ & $8.48\pm 0.14$ & $10 \pm 2$ (11)& 1670$\pm 550$ (6)& 666$\pm 450$ (9)& 0.4$\pm 0.08$\\
\hline                                   
\end{tabular}
\tablefoot{log $\epsilon$(X)$=$ log $n_x/n_H +12$, where log $n_X$
is the number density of element X. The number between parenthesis
indicates the number of lines used. The N and O abundances are adopted from the literature. We exclude features with both
a $^{17}$O and $^{18}$O contribution (see text).}
\end{table*}

For each of the model atmospheres adopted, synthetic LTE spectra
were calculated in the region 1850-2200 cm$^{-1}$ with a step of
0.0002 cm$^{-1}$ using the TURBOSPECTRUM v9.02 code described in
\citet{alv98}, and the line list given above. The theoretical
spectra were convolved with a Gaussian function with a FWHM$\sim
600-800$ m${\AA}$ to mimic the spectral resolution plus the
macroturbulence parameter. In order to estimate the C abundance
and the $^{12}$C/$^{13}$C ratio, we selected a number of
$^{12,13}$CO lines that are weak, unblended and, apparently, not
affected by the procedure for removing telluric absorption.  We
note that variations in the N abundance by $\sim\pm 0.3$ dex have
no effect in the synthetic spectrum. Also, the $^{12}$C$^{16}$O
lines are not very sensitive to changes to the O abundance by
$\sim\pm 0.2$ dex, thus we decided to adopt in our stars the N and
O abundances derived by \citet{ryd09} and
\citet{ram09}\footnote{CN lines are very weak in the 5 $\mu$m
region and cannot be used to derive the N abundance directly.}. It
is important to note that in the selection of these CO lines we
took into account the fundamental problem existing in the 5 $\mu$m
spectrum of K giants \citep[for example][]{hea78,ryd02,tsu09}: namely,
the fact that the CO fundamental lines cannot be interpreted with
a photospheric model only. These lines show an excess absorption
(lines with equivalent widths log $W/\nu > -4.75$) which is
probably non-photospheric in origin. Figures 1 and 2 show
clearly that the central cores of many CO lines cannot be
reproduced using a 1D photospheric model in a LTE calculation.
\citet{tsu08,tsu09} proposed instead that the extra-absorption
originates from cool molecular layers referred to as a
quasi-static molecular dissociation zone; this is sometimes named
MOLsphere. Indeed, the formation of these molecular clouds in the
outer atmosphere appears to be a basic feature of all the red
giant stars from early-K to late-M types \citep[see][for more
details]{tsu09}. Therefore, our estimates of the C absolute
abundance and of the C and O isotopic ratios are based on a
careful selection of the CO lines (i.e. log $W/\nu\leq -4.75$), so
that in principle the approximation of LTE synthetic spectra using
canonical 1D photospheric models should be valid. By considering
this, we derive in Arcturus a C abundance in agreement with that
in \citet{ryd09}. In Aldebaran we derive a C abundance larger by
0.15 dex than the one by \citet{ram09} (see Table \ref{tabobs}).
On the other hand, the carbon isotopic ratios derived in both
stars agree also with previous estimates, being 9 and 10 for
Arcturus and Aldebaran, respectively (see Table \ref{tabobs}).

\begin{figure*} \centering
\includegraphics[width=15cm]{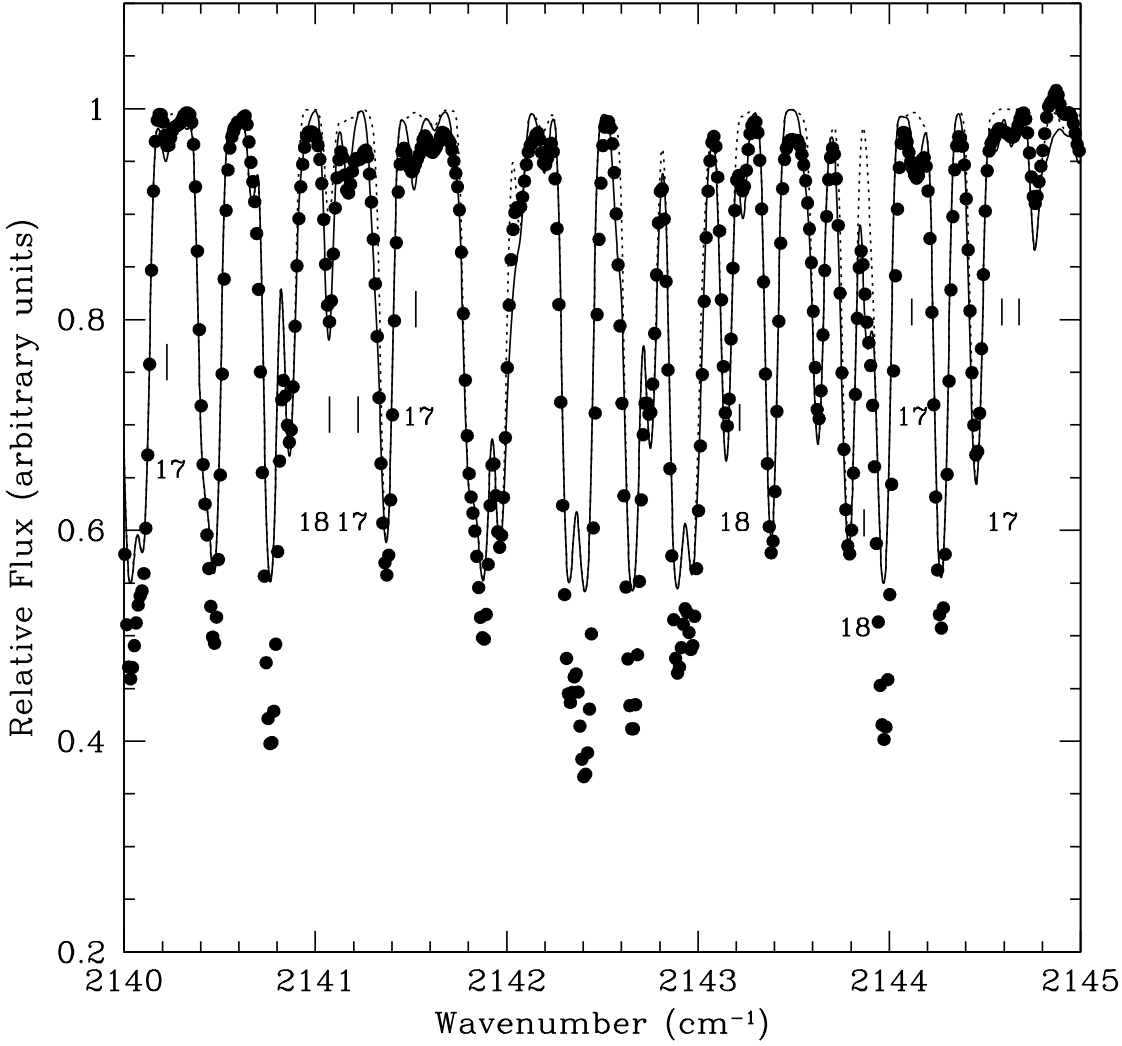}
\caption{Comparison of observed (dots) and synthesized spectra
(lines) of Arcturus for different O ratios. Dotted line:
$^{12}$C/$^{13}$C $=9$ and $^{17}$O/$^{16}$O $=^{18}$O/$^{16}$O
$=0$; continuos line: the same $^{12}$C/$^{13}$C with
$^{17}$O/$^{16}$O $=3030$, $^{18}$O/$^{16}$O $=1660$. Note that
the cores of the more intense CO lines are not reproduced by the
theoretical spectrum (see text). Some of the C$^{17}$O and
C$^{18}$O lines used are marked.} \label{fRGB1}
\end{figure*}

Once the CNO abundances and the C isotopic ratio were estimated,
in each star the absorption features due to $^{17}$O and $^{18}$O
were fitted by the synthetic spectrum varying the abundance of
these isotopes in order to give the best fit to each feature, one
at a time. We selected very carefully these lines to avoid as much
as possible blending, cases where the position of the continuum
was regarded as uncertain and/or the presence of weak telluric
lines in the rationed spectrum. This resulted in a lower number of
useful $^{17}$O and $^{18}$O lines as compared to the study by
\citet{har84}, performed in the same spectral region. However,
contrary to these authors, we did not assign any weight to any
feature to compute the final O ratios. We excluded also features
where both $^{17}$O and $^{18}$O were contributing. The abundances
derived from the various features selected in this way were then
combined to give a mean (Table 3).

The major source of uncertainty in the derivation of the C and O
ratios is the dispersion in the ratios obtained from the different
lines. Uncertainties due to errors in the atmospheric parameters
(see Table \ref{phys}) are minor as compared to these. As
mentioned before, changes in the metallicity of the model
atmosphere up to $\pm 0.25$ dex has no impact in the derived
ratios. The same is true with changes in the N abundance by $\pm
0.3$ dex and/or $\Delta$ T$_{eff}=\pm 100$ K. Uncertainties in the
gravity and microturbulence of the order of those quoted in Table
\ref{phys}, imply errors in the O ratios not exceeding about $\pm
100$ in both stars. A larger impact on the final error comes from
the uncertainty in the C and O abundances and the
$^{12}$C/$^{13}$C ratio. All these sources of error, once added
quadratically, give a total uncertainty of $\pm 180$ for Arcturus,
and $\pm 230$ for Aldebaran. These figures may be safely applied
to both O ratios. In Table \ref{tabobs} we indicate the total
error in the C and O ratios after including the dispersion in the ratios
between different lines. It is evident that the dispersion from
different lines accounts for most of the total error. Systematic
errors may be present, like for example the uncertainty in the continuum
position and departures from LTE. Due to the weakness of the
$^{17}$O and $^{18}$O lines used, errors in the continuum position
should affect the $^{17}$O and $^{18}$O abundances almost equally,
so that the ratio $^{17}$O/$^{18}$O is probably more reliable. For
the same reason, departures from LTE should be small in the layers
where the key features are formed.

When comparing our oxygen ratios with those derived in
\citet{har84}, we agree within the error bars only in the
$^{16}$O/$^{18}$O ratio in Aldebaran; we derive considerably
larger  $^{16}$O/$^{17}$O and $^{17}$O/$^{18}$O ratios in both
stars. {\bf  The differences in the atmosphere parameters adopted
(T$_{eff}$, log g, $\xi$) are not strong enough to explain the
discrepancies. In fact by using atmosphere models from the Gustafsson et al. (2008)
grid with the same stellar parameters and CNO abundances as adopted by \citet{har84}\footnote{{\bf These authors
give only the C abundance derived ([C/H]$=-0.7$ and $-0.3$, respectively) without any indication of
the N and O abundances derived/adopted nor error bars. Thus, in this test we scale the N and O abundances according to the metallicity
given in \citet{har84} respect to the solar abundances from \citet{lam78} (we infer that these solar
abundances were adopted by these authors; note this circumstance in figures 5 and 7 below, where we adopted
a conservative error bar of $\pm 0.2$ dex in the CNO abundances by Harris \& Lambert)}} we obtain:
$^{16}$O/$^{17}$O$=2325$ and $^{17}$O/$^{18}$O$=1430$ for Arcturus, and 
$^{16}$O/$^{17}$O$=1540$ and $^{17}$O/$^{18}$O$=560$ for Aldebaran. The ratios
are reduced significantly but not enough (note that we still agree in the $^{16}$O/$^{18}$O ratio in Aldebaran).
We recall that \citet{har84} used atmosphere models from \citet{bell76} and
\citet{joh80}. The \citet{bell76}'s models are in fact the ancestors of the
new grid of spherical MARCS atmosphere models by Gustafsson et al. (2008). These new models
considerably improve the atomic and (mainly) molecular opacity treatment, in particular
for giant stars, as well as many other physical approximations (see Gustafsson et al. 2008 for
a detailed discussion), so that we consider that new MARCS models mimic much better
the real atmospheres of giants than the original ones by Bell et al. (1976) and 
Johnson et al. (1980).
An additional source of discrepancy which is very probably the main cause of our
different findings is the line list used for the CO molecule and the
method employed for the continuum placement. First, we note that \citet{har84} only include
CO lines in their synthetic spectra computations, while we consider a number of molecular and
atomic species (see previous section) in the 4.5 $\mu$m spectral region. Indeed, the CO molecule
dominates the absorption in this spectral range, but we checked that the remaining 
molecular species (mainly CN, C$_2$ and
OH) introduce a {\it veil} of absorption which increases the line intensities and thus affects
the isotopic ratios derived. In particular the $^{16}$O/$^{17}$O and $^{16}$O/$^{18}$O ratios increase.
Secondly, the {\it gf-}values of the CO lines used by \citet{har84} were obtained from \citet{cha83}, while
here we used those by Goorvitch (1994) on the basis of a more up-to-date electric dipole moment
function. As a consequence, Goortvitch (1994) indeed reported differences up to $3\%$ in the {\it A}-values with respect
to \citet{cha83} and up to a factor $\sim 6$ in the dipole moments in some of the CO isotopes. On the other hand, note that the
isotopic lines used are very weak so that the choice of the specific lines might affect the results, in a systematic 
manner (see above the discussion on the selection of the lines) if there is a
systematic difference in the continuum placement. Definitely, the discussion above might explain the differences between the
O ratios derived here and those in \citet{har84}.  We believe that our figures are more reliable because they
are based on the use of more accurate stellar parameters and CNO abundances, better atmosphere models and more 
complete and accurate line lists.  
}
\begin{figure}
\centering
\includegraphics[width=\columnwidth]{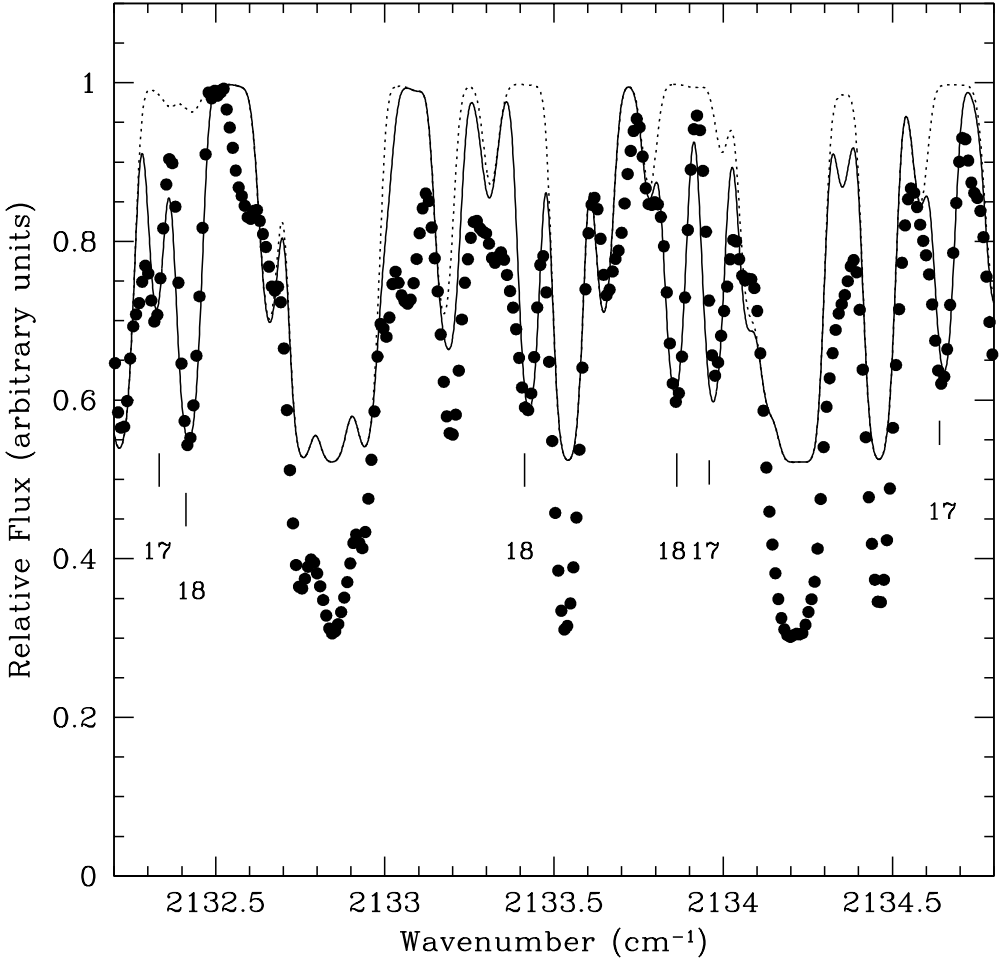}
\caption{As Figure \ref{fRGB1} for Aldebaran (filled circles) in
another spectral region. Dotted line: $^{12}$C/$^{13}$C $=9$ and
$^{17}$O/$^{16}$O $=^{18}$O/$^{16}$O $=0$; continuous line: the
same $^{12}$C/$^{13}$C with $^{17}$O/$^{16}$O $=3030$,
$^{18}$O/$^{16}$O $=1660$. Note again the difficulty to fit the
cores of the strongest CO lines (see text). Some of the C$^{17}$O
and C$^{18}$O lines used are marked.} \label{fRGB22}
\end{figure}

\subsection{Physical stellar parameters} \label{sec24}

In order to check the reliability of the mass estimates available
in the literature (see Table \ref{phys}), we used the FRANEC code
to construct our theoretical HR diagrams. (We warn the reader that the
theoretical mass estimate for a single stellar object, obtained by fitting its
HR diagram, is quite uncertain. On the other hand, more robust methods, such as the
classical isochrones fitting, cannot be applied to our stars). This was done in the
following way. i) First we selected the range of initial masses
$M_{ini}$ for which the theoretical HR diagrams could fit the $L,
T_{eff}$ values measured and the available estimates for the
stellar radii of our stars within their uncertainties. ii) Then,
we computed, over the mass range found, a grid of extra-mixing
models, looking for the combination of $M_{ini}$ values and
transport parameters that allowed the best global fit.

With this procedure, we found very easy to reproduce the
observations for $\alpha$ Tau, assuming initially solar abundance
ratios (see section \ref{sec3}); our best estimate for the mass
agrees very well with previous determinations. For Arcturus,
instead, things were considerably more complex. This is a slightly
metal-poor star for which guessing the initial abundance ratios is
not straightforward. For C and N we had to rely on the literature,
as the measured data are certainly not the initial ones, being
modified by FDU and extra-mixing. However, there are suggestions
that Arcturus belongs to a peculiar {\it stream} of stars in the
vicinity of the Sun; some of them date back to the early seventies
\citep{eggen}. Today this stream is considered as being the relic
either of an old dissolved cluster \citep{ram11} or of a captured
and disrupted dwarf spheroidal galaxy \citep{navarro}. Its
abundances seem to be rather similar to those of the Galactic
thick disk, but cautions are mandatory. Moreover, there is a
considerable dispersion of initial abundances in thick disc stars
at the metallicity of $\alpha$ Boo, so that we can only put weak
constraints on them. The initial values adopted here include a C
enhancement by $+$0.2 dex and a N under-abundance by 0.1 dex
\citep{bensby,matteucci}. With these choices, and using an initial
enhancement of $\alpha$-rich elements of +0.4 (including oxygen)
from \citet{pet93}, the model HR diagrams that can fit the
($L,T_{eff}$) data within the uncertainties correspond to a mass
interval from 1 to about 1.25 $M_{\odot}$. The final value we
adopted (1.2 $M_{\odot}$) is the only one for which extra-mixing
models can reproduce within the errors all the abundance
information discussed in Section \ref{sec23}. The uncertainties on
the initial C, N data make however our solution less robust than
for $\alpha$ Tau. Note that our value is slightly larger than the
one derived by \citealt{ram11} (1.08 $M_{\odot}$), who used as
reference the Yale code, with a lower $\alpha$ enhancement.

%

\section {Reproducing the abundances through an extra-mixing model}
\label{sec3}
Whatever the mechanism is that drives non-convective
mixing in red giants, \citet{Wasserburg1995} showed that it can be
approximated by a circulation occurring at a rate $\dot M$,
reaching down to a maximum temperature $T_{P}$, close to, but
lower than, the H-burning shell temperature. In a diffusive
approach the parameters would instead be the diffusion coefficient
$D$ and the total mass involved: the two approaches can be shown
to be roughly equivalent \citep{nol03} in most cases. One has however to
notice that, while this is certainly the case when the mixing speed
is not relevant (so that the abundance changes depend only on a path
integral of reaction rates), things might be different when the velocity of mixing
becomes an important issue, as diffusive processes are always slow, while
other transport mechanisms might not be so. In fact: i) the time available for
mixing is not infinite; and ii) the nuclei involved are sometimes unstable
with a relatively short half-life: se for example the case of $^7$Be, decaying into $^7$Li
\citep{pal11x}. On similar grounds, recently serious doubts have been
advanced on some of the proposed mechanisms, like rotation and
thermohaline mixing \citep{cl10,pal11x} because of the small
diffusion coefficients (or, alternatively, the slow mixing speeds) they can provide, which might 
make them inadequate to yield the observed abundance changes in the finite time assigned by 
the duration of the evolutionary stage \citep{denimay}.

An important input to the models is the initial CNO admixture of
the stellar composition. While for $\alpha$ Tau, a typical
thin-disk red giant of relatively high metallicity, we can safely
assume solar elemental ratios, as we already mentioned for $\alpha$ Boo this is certainly
not the case (see discussion in Section \ref{sec24}). In our
procedure, once the initial CNO abundances are selected, they are
employed for choosing the opacity tables to be used. Enhancements
in $\alpha$ elements and, to a lesser extent, in C easily
introduce large changes in the models, from both the points of view of
both nuclear physics (CNO burning efficiency) and radiative transfer (opacities). We
underline this important point because the theoretical HR diagram
and the ensuing mass estimate strongly depend on that. Note that
using the most recent set of $alpha-$enhanced opacities \citep[see for example][]{ferg05}
theoretical curves are moved to redder regions of the HR diagram than previously
found. Thus, we warn that all the mass estimates available so far in the
literature are actually much more uncertain than currently supposed
\citep[see for example][]{ver05,tsu09,ram11}.

\begin{figure}
\centering{
\includegraphics[width=\columnwidth]{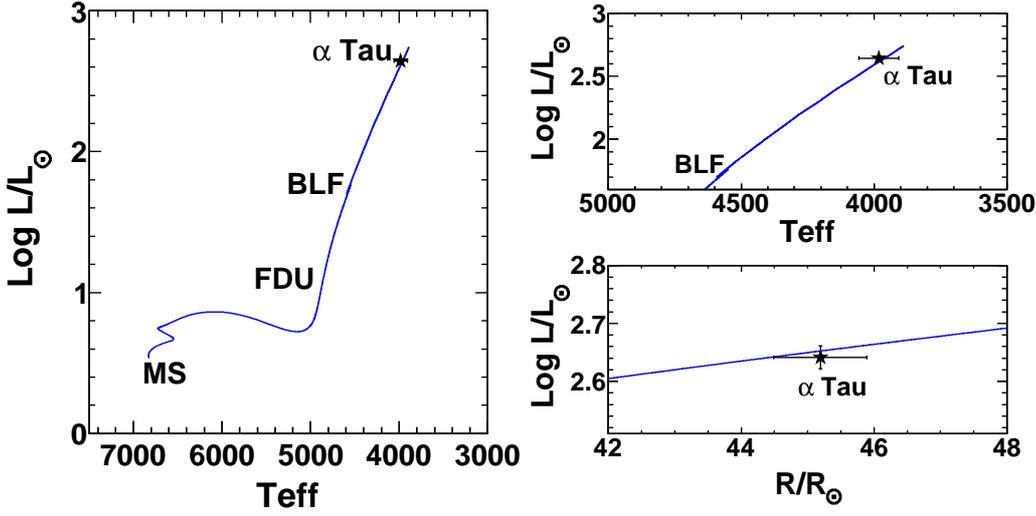}
\caption{Left panel: a comparison of the observed $L$ and $T_{eff}$
values of $\alpha$ Tau with the adopted evolutionary track in the HR
diagram, as derived from the FRANEC code. Right panel: zoom of the
previous plot (top) and a comparison of model and observations for the
$L, R_{star}$ relation. {\bf The main sequence (MS), first dredge-up (FDU) and (BLF) positions are marked.}} }
\label{fRGB3}
\end{figure}

Our best solution implies, for $\alpha$ Tau $M = 1.3 M_{\odot}$, [Fe/H] =
$- 0.15$; for $\alpha$ Boo, $M = 1.2 M_{\odot}$,
[Fe/H] = $- 0.5$, [$\alpha$/Fe]$=0.4$, [C/Fe]$=0.2$.
As mentioned before, we performed a grid of calculations for both
stars: for each choice of the mass allowing a fit (within the
uncertainties) to the $L, T_{eff}$ data, we derived the
extra-mixing parameters trying to reproduce the chemical
abundances; finally, we adopted the case allowing the best
compromise in the fit of all the available data. There are no
ambiguities on $\alpha$ Tau with this procedure, and the resulting
HR diagram is shown in Figure \ref{fRGB3}.

For $\alpha$ Boo, instead, the solution that we finally adopt
appears to require a mass higher than so far assumed and is
strongly dependent on the initial CNO assumed. This point is
further commented later (Section \ref{para}). The results for the
HR diagram and radius of $\alpha$ Boo are displayed in Figure 4.

\begin{figure} \label{fRGB4} \centering{
\includegraphics[width=\columnwidth]{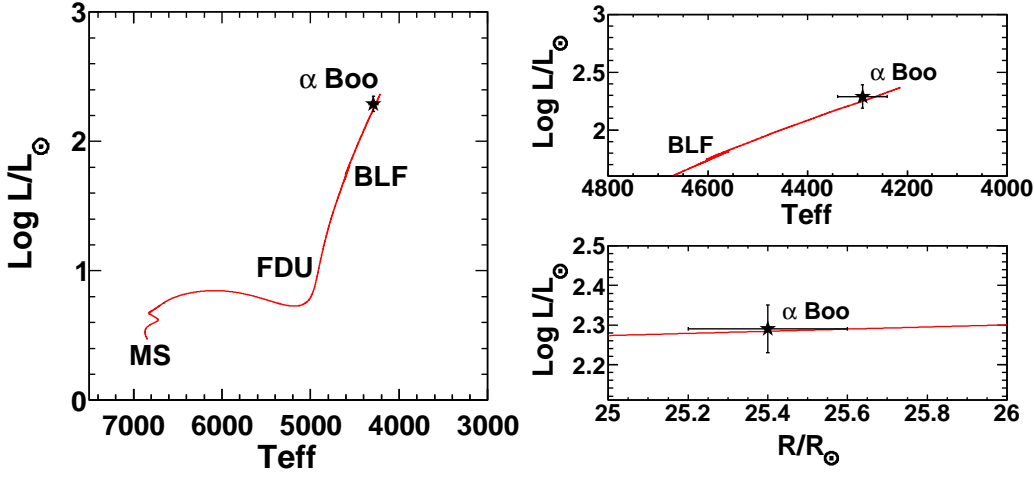} \caption{The
same as Figure \ref{fRGB3} for $\alpha$Boo. Note the peculiar
initial CNO content for this star (see text).}}
\end{figure}

According to the above discussion, from the estimate of the mass
we derive the time spent on the RGB from the bump in the luminosity
function (the small dent in each RGB track indicated by the label
"BLF") to the moment in which
the observed values of $L, T_{eff}$ are attained. This is the time
available for extra-mixing to operate: it turns out to be 46 Myr
for $\alpha$ Tau and 37 Myr for $\alpha$ Boo.  This knowledge
allowed us to determine the mixing parameters on observational
grounds (albeit with the mentioned cautions for $\alpha$ Boo).

\begin{figure*}[t]
\centering{\includegraphics[width=11cm]{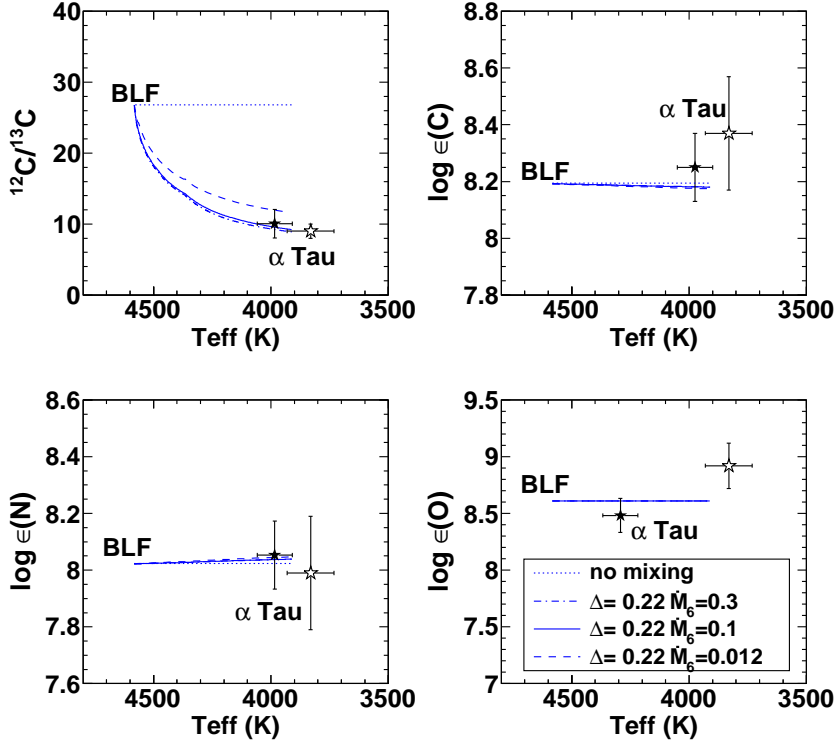}
\caption{The evolution of the $^{12}$C/$^{13}$C ratio and of the
elemental CNO abundances as a function of the model effective
temperature in selected extra-mixing runs performed for $\alpha$
Tau, as compared to observed data. Black stars show the observational data presented in this work,
{\bf while open stars report abundances from \citet{har84}}.
A case without extra-mixing is reported for comparison (dotted line). See text for details.}
}
\label{fRGB5}
\end{figure*}

\begin{figure*}[t]
\centering{\includegraphics[width=11cm]{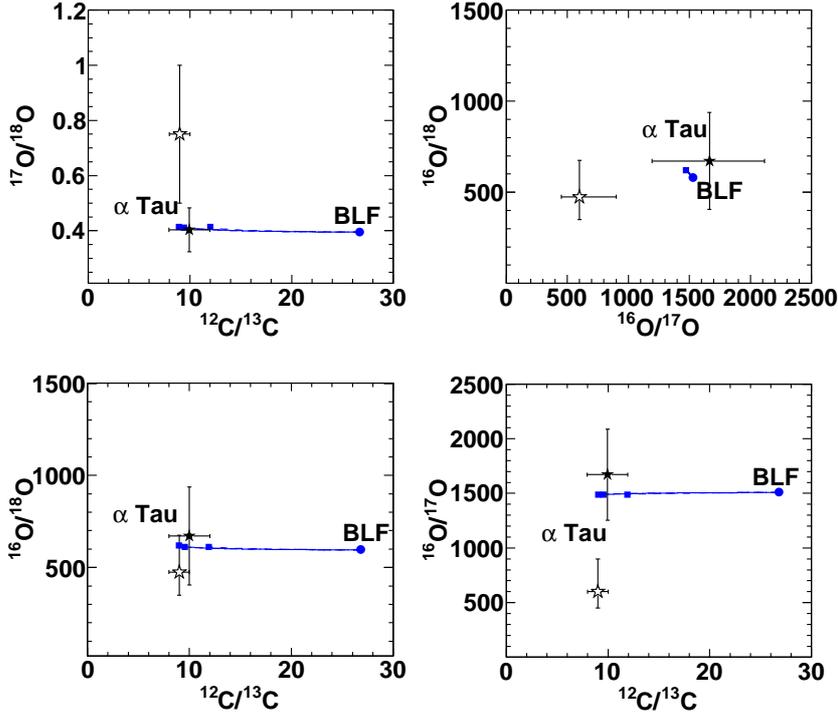}
\caption{Combinations of isotopic C and O ratios for selected
extra-mixing runs performed for $\alpha$ Tau and for a case
without extra-mixing (blue spot), as compared to observed data. The models and
the meaning of the line types are the same as in Figure 5.
The small squares indicate the envelope isotopic mix reached by each mixing
case when the model effective temperature is comparable with the observed one. At the same
mixing depth, the higher the mixing rate $\dot{M}_6$, the lower the reached value of the carbon isotopic ratio the lower is the value reached for the carbon isotopic
ratio. {\bf Open stars are the isotopic ratios derived by \citet{har84}.}}} \label{fRGB6}
\end{figure*}

\subsection{The technique of the computations}

In order to fix
quantitatively the parameters of extra-mixing using as constraints
the observations of $\alpha$ Boo and $\alpha$ Tau discussed so
far, we adopted the formalism by \citet{nol03}, and made therefore
parameterized calculations. After deducing the parameters that
allow us to fit the measured abundances, we derived the
corresponding mixing speeds needed to achieve the observed abundances
in the assigned time. On this basis, we could analyze which
of the processes proposed so far in the literature offers a
plausible physical mechanism for driving the mixing.

In our procedure we adopted a post-processing code to compute the
coupled phenomena of H-burning and transport, taking the detailed
stellar parameters from the output of the stellar evolution code
\citep[FRANEC: see][]{cris11}, which provides us with the physical
structure of the star.

For describing the nuclear physics phenomena coupled with dynamics one
can simply use the total derivatives of stellar abundances:
\begin{equation}
\label{eq2}
\frac{dN_i}{dt}=\frac{\partial{N_i}}{\partial{t}}+\frac{\partial{N_i}}
{\partial{M}}\frac{\partial{M}}{\partial{R}}\frac{\partial{R}}{\partial{t}}
\end{equation} where the partial time derivative due to
nucleosynthesis is:
\begin{equation}
\label{eq1}
\frac{\partial{N_i}}{\partial{t}}=-N_{p}N_{i}\lambda_{i,p}+N_{p}N_{i-1}\lambda_{i-1,p}-N_{i}\lambda_{d}+N_{i'}\lambda_{d}
\end{equation} and the second term is due to mixing. Here the
parameters $\lambda_{i,p}$ are the reaction rates and $\lambda_d$ are
the decay rates. For the nuclear parameters we adopted the recent
upgrades suggested by \citet{ade11} and \citet{ili10}.  Details on our
technique for computing extra-mixing and on the descending numerical
computations are presented by \citet{pal11y,pal11x}.

\begin{figure*}[t]
\centering{\includegraphics[width=11cm]{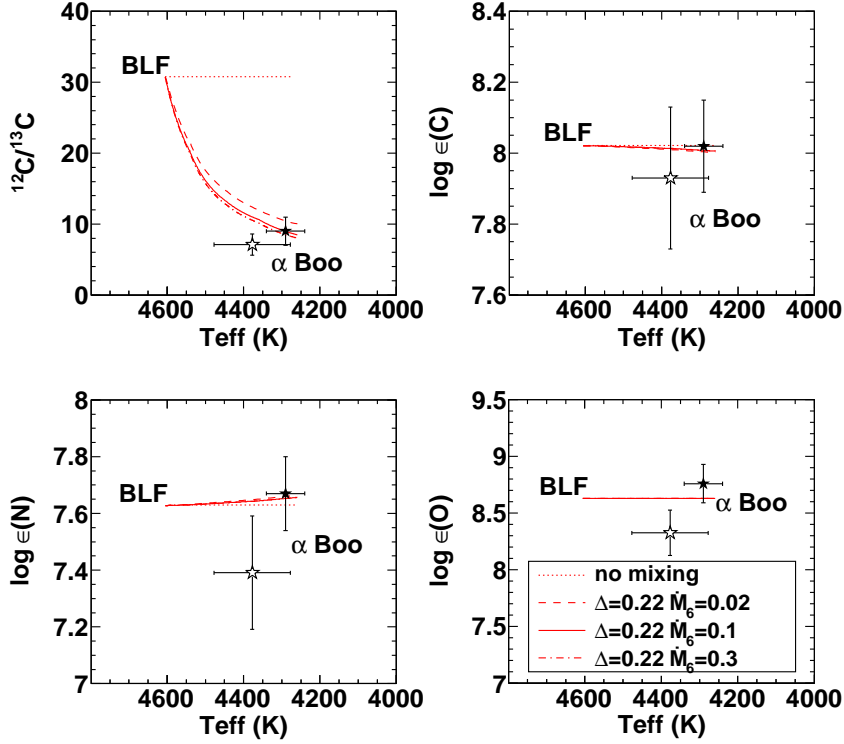}
      \caption{The same as Figure 5 for $\alpha$ Boo.}}
 \label{fRGB7}

\end{figure*}

\begin{figure*} [t!!]
\centering{\includegraphics[width=11cm]{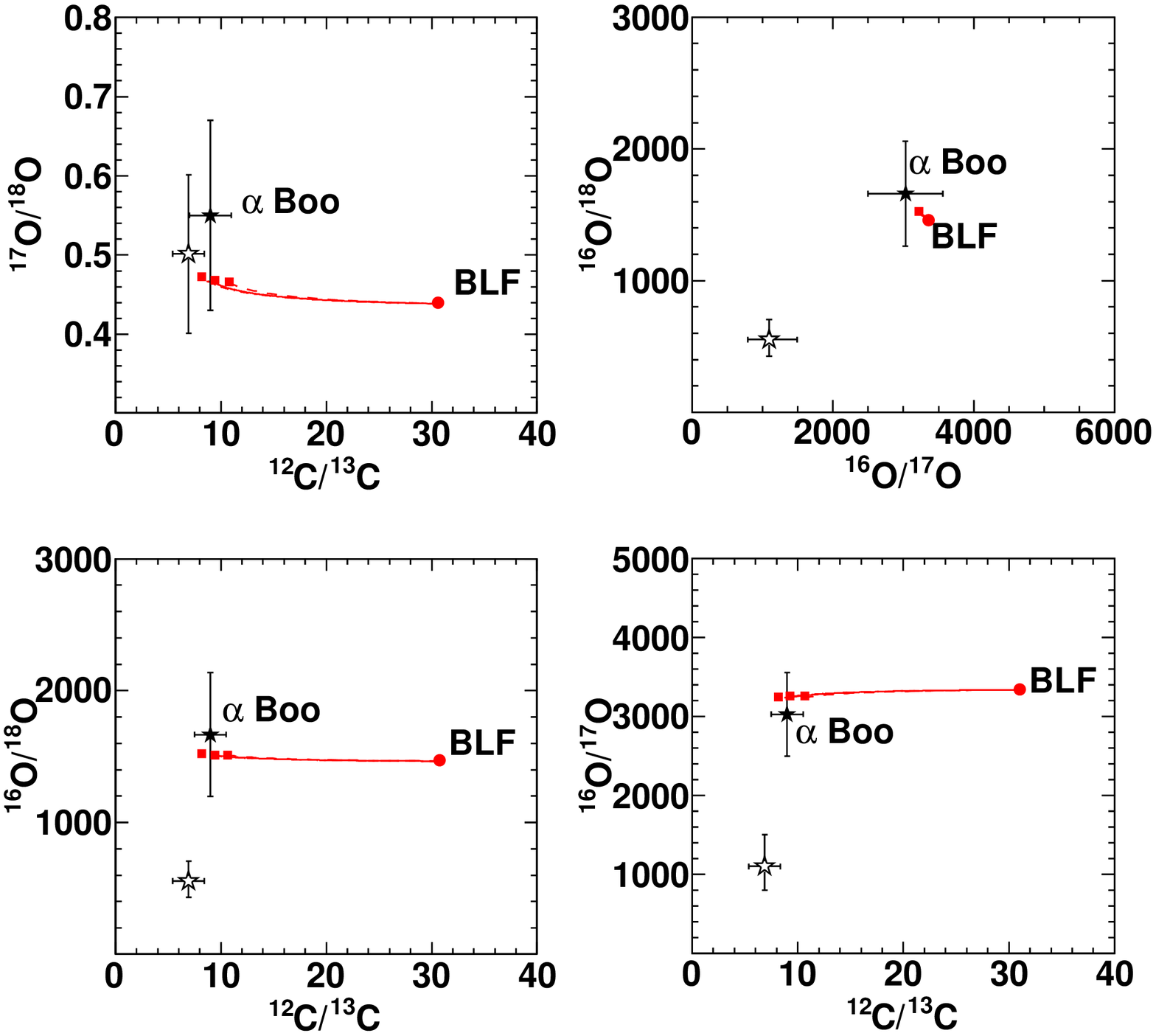}
      \caption{The same as Figure 6 for $\alpha$ Boo.}}
         \label{fRGB8}
\end{figure*}

\subsection{The parametric results}\label{para}

\begin{figure*}[t!!]
\centering{
\includegraphics[width=\columnwidth]{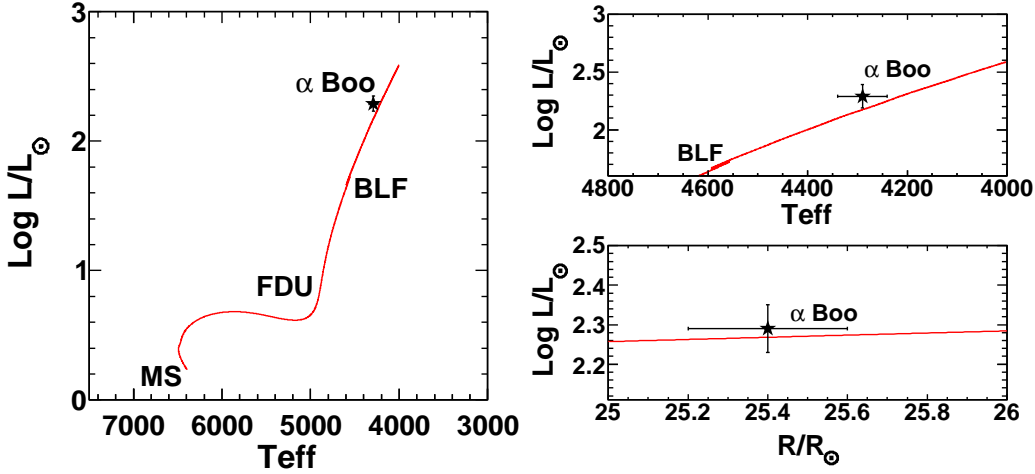}
\caption{The same as Figure 4, but with a choice of $1.08~M_{\odot}$  (see
text for details).}}
\end{figure*}

Leaving as free parameters the circulation rate $\dot M$ and the
temperature $T_P$ of the deepest layers reached by the non-convective
mixing, we could profit of previous work done for a wide sample of RGB
stars by \citet{pal11y,pal11x} for limiting the interval of variation
of these parameters. Following the formalism used by the above
authors, the circulation rate was expressed in units of 10$^{-6}$ $M_{\odot}$/yr,
through the parameter $\dot M_6$.  Concerning the temperature of the deepest layers attained by the mass transport, $T_P$, this was considered through the parameter $\Delta = \log T_H - \log T_P$, $T_H$ being the temperature at which the maximum energy from H burning in the shell is released. Although this is certainly a not-very-intuitive way of expressing the mixing depth, it offers a sort of rule-of-thumb criterion, established by \citet{nol03}. If $\Delta$ is lower than about 0.1, post-process mixing models are rather safe, in the sense that any
nucleosynthesis occurring during the transport will not add significant energy
to the stellar budget, thus not altering the reference stellar structure.

\begin{figure*}[t!!]
\centering{
\includegraphics[width=0.8\columnwidth]{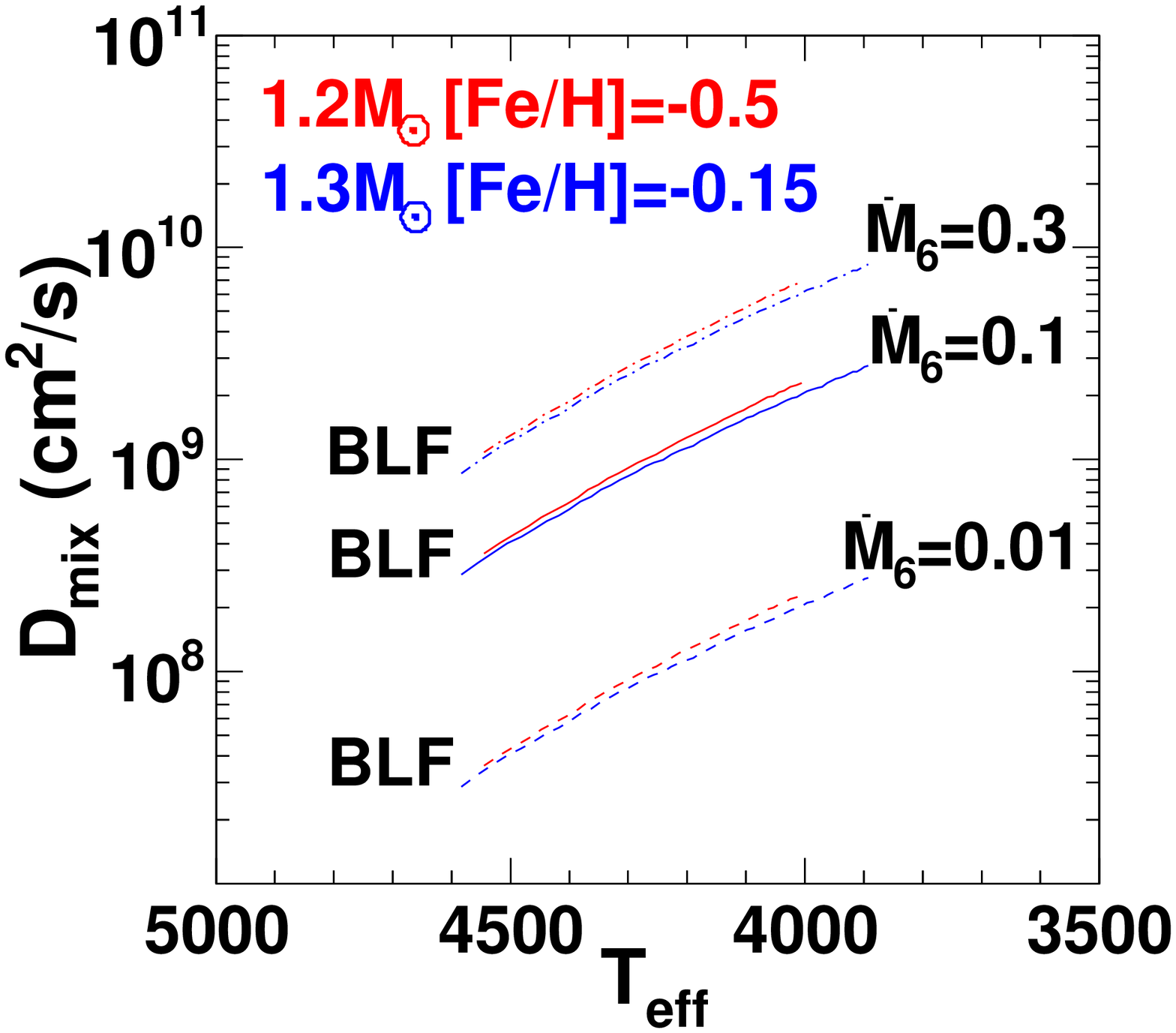}
\caption{Diffusion coefficients approximating our circulation
models under the hypothesis of a diffusive treatment (see text for
details). The parameters of the considered cases are shown in the
labels.} }
\end{figure*}

On this basis $\dot M_6$ was allowed to vary in the range from
0.01 to 0.3 and $\Delta$ in the range from 0.18 to 0.22. For each
combination of values of the two parameters we computed the
corresponding post-process mixing models, starting from the
stellar structure along the RGB as provided by full calculations
(with no extra-mixing) made with the FRANEC code. The technique
adopted is that of reading from the outputs of the stellar code
the physical parameters at and above the H burning shell, up to
the convective envelope, using them to compute the outcomes of the
coupled processes of burning and mixing with the parameters
adopted; the inputs from the full model stellar structure are
refreshed periodically, in order to be sure to maintain coherence
between the extra-mixing calculations and the real physics of the
star.

Among the many
runs performed we show in Table 4: i)  those computed for $\alpha$ Tau,
using our best choice for its mass; and ii) those computed for $\alpha$ Boo,
using our mass estimate (1.2 $M_{\odot}$)  and iii) those run by
adopting the mass values given by
\citet{ram11}.

The comparison of model sequences for isotopic and elemental abundances with
observed data is presented in Figures from 5 to 8 for the best cases
selected from Table 4. Here the curves with different line types refer to
different choices of the parameters, according to what is explained in the labels.

In particular, Figure 5 shows the evolution of the
$^{12}$C/$^{13}$C ratio and of the elemental CNO abundances along
the RGB for extra-mixing models adopted for $\alpha$ Tau, as a
function of the effective temperature (which in this case is a
proxy for time, as red giants become progressively redder and
cooler when they evolve). The label BLF identifies the abundances
as determined by the FDU (these are the same as those
characterizing the envelope layers at the Bump of the Luminosity Function).
It is clear from the plot that there is no way of
explaining the carbon isotopic ratio with a model that does not invoke
extra-mixing. Such a case (dotted
curve) presents in fact a too large $^{12}$C/$^{13}$C ratio when
compared to the observations. The same conclusion can be
obtained by an inspection of Figure 6, where the model
combinations of the C and O isotopic ratios are compared with the
observations. In this plot the case without extra-mixing is
represented by a single dot, since neither carbon nor oxygen
change their isotopic ratios along the RGB. An extra-mixing model with
a too deep penetration (those with $\Delta$=0.18) can be excluded
on the basis of its large $^{17}$O/$^{18}$O ratios and of its low
absolute carbon abundance (see Table \ref{tabtot}). A case with an intermediate
depth ($\Delta$=0.20) would alleviate the problem of oxygen isotopic ratios,
but would still predict a too low carbon abundance. Thus, for $\alpha$ Tau a
rather shallow extra-mixing is needed ($\Delta$=0.22). Concerning
its efficiency in terms of mass circulation, all the cases lie within the
observed range, apart from that with the lowest efficiency ($\dot M_6=0.01$).
The $\dot M_6=0.012$ case shows the maximum allowed $^{12}$C/$^{13}$C ratio, while
extra-mixing models in the range $0.03\leq\dot M_6 \leq0.3$
reproduce well all the observed constraints. We remind again that the
most critical point is certainly the $^{12}$C/$^{13}$C ratio, even
if oxygen isotopic ratios help in constraining the extra-mixing
depth. The choice of the parameters found to be appropriate is
in good agreement with the cases run by \citet{pal11y,pal11x}, so
that $\alpha$ Tau appears to be really a typical template for
Population I red giants, also from what concerns the mixing
processes.

Figures 7 and 8 integrate the previous picture by showing the more
complex case of $\alpha$ Boo. As discussed previously, here the
constraints from elemental abundances are rather weak, depending
on assumptions made for their unknown initial CNO values. Despite
this caution, it is noticeable that the extra-mixing parameters
determined for $\alpha$ Tau ($0.03\leq \dot M_6 \leq 0.3$, $\Delta =
0.22$) also reproduce well the constraints for $\alpha$ Boo. This
is, in our opinion, a very interesting finding, as the two red
giants have a different metallicity. Many of the computed cases
have been excluded on the basis of reasonings similar to those exposed
for $\alpha$ Tau (see Table \ref{tabtot}).
Considering the full extension of  the error
bars for the observations, extra-mixing cases with a rather slow
circulation ($0.01\leq \dot M_6 \leq 0.03$) cannot be a priori excluded for $\alpha$ Boo.
Notwithstanding, the global quality of the fits is much better with larger
$\dot M_6$ values, i.e. with a choice similar to that for $\alpha$ Tau.

From a further inspection of Table \ref{tabtot} it emerges that, as far as
the abundances are concerned, adopting
a slightly lower mass than found by us (for example for 1.08 $M_{\odot}$, the
value determined by \citealt{ram11})
several (albeit not all) data can be reproduced
with a different choice of the parameters ($\Delta \sim$
0.18, $\dot M_6 \sim 0.015$), that is with a slightly deeper and slower
extra-mixing. Remarkable discrepancies emerge in this case only
for the absolute elemental abundances, whose modeling suffers for
the uncertainties in the initial composition already discussed.
However, in the case with $M$ = 1.08 $M_{\odot}$ a
poorer fit to the luminosity and radius of $\alpha$ Boo is obtained
(see Figure 9).  We therefore maintain our previous choice of the mass
(1.2 $M_{\odot}$) as our best case, but we remark that it, too, is
rather uncertain.

\section{Discussion and conclusions}

We notice that the values of the extra-mixing parameters deduced
for our stars nicely fit with those previously found to be typical
for red giants of Population I, from their CNO isotopic and
elemental abundance ratios as well as from their Li content
\citep{pal11y,pal11x}. Under these premises, we are tempted to
conclude that extra-mixing during the RGB phase presents common
properties for solar or moderately low metallicities.
The idea is suggestive, although we
cannot at this level draw too firm conclusions. A wider
sample of moderately metal-poor, well measured red giants,
permitting reliable statistics \citep[like the one selected
by][for higher metallicities]{pal11y} would be required for this.

It is relevant to compare our findings to those proposed in the
literature, adopting the complementary view of a diffusive
approach. In particular, \citet{den10} and \citet{denimay},
recently performed 1D studies of the effects of thermohaline
diffusion (which can be easily compared to our discussion) and
then substantiated their results with 2D- and 3D-simulations. In
\citet{den10} it was shown that thermohaline mixing might
guarantee diffusion coefficients $D_{mix}$ smaller than a few
10$^6$ cm$^2$/sec (see especially their Figure 3, in which
$D_{mix}$ was normalized to the the thermal diffusivity. This last
was assumed to be of the order of 10$^8$, as specified in Table 1
of that paper). However, in the subsequent discussion (in
particular from Figures 7, 11 and 12) the author makes clear that,
in order to fit all the red giant data (including the values of
C/Fe and the isotopic ratios $^{12}$C/$^{13}$C lower than about
15) one would need much larger values of $D_{mix}$. These last
seem however to be compatible with thermohaline mixing only for
low metallicities. In our case we can make a rough estimate of
$D_{mix}$ from the simplified correspondence between the
circulation and diffusion treatments established by \citet{nol03},
i.e. $D_{equiv} \simeq (l \times \dot M)/4 \pi \rho r^2$.

For $\dot M$ = (0.3 $-$ 3)$\times$10$^{-7} M_{\odot}$/yr (the
range found to be good for our stars) and adopting the values of
the other parameters from the stellar code outputs we obtain
values of $D_{equiv}$ as those plotted in Figure 10 (see also the
last column in Table \ref{tabtot} where average $D_{equiv}$ values
of each extra-mixing case are reported). For the acceptable cases they
cover a range centered around a few 10$^9$ cm$^2$/sec (this last
value being roughly the average).

We can compare this with the data of Figure 12 in \citet{den10},
where the curve for $D_{mix}$ providing the best fit to the data
is plotted as a function of the radius. The radius of the
convective envelope border in $\alpha$ Tau, in the phases after
the BLF, spans the range $log[r/R_{\odot}]$ = $-$0.13 to
$-$0.07. For these values the average value $D_{mix}$ is again
$10^9$ cm$^2$/sec, in agreement with the value found by us. Hence,
with a completely independent treatment, we confirm the results of
the quoted paper. The diffusion coefficient required for
explaining the observations of $\alpha$ Tau (and the more
uncertain $\alpha$ Boo) must be quite large (Such a value for $D_{mix}$ implies in our
cases a velocity of a few hundredths cm/sec).

Note that also the treatment by \citet{pal11y}, favoring $\dot M$ values
very similar to those of $\alpha$ Tau and $\alpha$ Boo, would provide the same
consequences for the values of $D_{mix}$.

By pursuing a similar discussion in the framework of multi-D models of
the radiative zones \citet{denimay} showed that this is related to the
aspect ratios (length over diameter) of the dynamical instabilities
generated in the simulations: with thermohaline mixing these aspect
ratios would be too small, i.e. the unstable blobs would be too
similar to "bubbles" instead of the required "finger"-like structures.

Whatever approach is used, the results seem to converge in saying
that pure thermohaline diffusion might have difficulties in explaining the observed
abundances of high-metallicity red giants, at least when it is taken
alone; the possibility of a modified magneto-thermohaline mixing was
envisaged by \citet{den09} but not yet substantiated by detailed
models. When also the results by \citet{cl10} are considered, one sees
that any diffusion induced by rotational effects is in its turn
insufficient; in this case, actually, $D$ is too small by several
orders of magnitudes (see Figure 9 in that paper).

{\bf We have to notice that in some recent works \citep[see for example][]{angel} the
authors wisely avoid the use of the term "thermohaline diffusion" for indicating
chemical readjustments started by a molecular weight inversion; they prefer the name
{\it $\Delta \mu$ mixing}. The use of the term diffusion could have remarkable implications.
Indeed, while diffusion is intrinsically a slow phenomenon and thermal diffusion
should in fact be slow, it is not
guaranteed that, in presence of an inversion of the $\mu$ gradient, the chemicals would diffuse
with a speed comparable to that of heat. We have no elements at this stage
to exclude that non-diffusive mixing of a suitable velocity might occur as a consequence
of a  "$\Delta \mu$" effect: in that case, it would offer a realistic mechanism.
Our analysis only underlines that very slow mixing, as in diffusive processes, would be
inadequate for explaining the chemical abundances; but there is clearly a lot we have still
to learn about the real physical processes.}

One has to underline that magnetic buoyancy, recently advocated in
various ways by \citet{bus07,Busso2008,den09} is suitable to provide
the mixing velocities (or the $D_{mix}$ values) we require for the
two stars examined. Looking
back at old, seminal works by E.N. Parker \citep[for example][]{par74} one
sees that the velocity of buoyant magnetic structures, in presence of
thermal exchanges with the environment, is roughly $v = K/a^2$, where
$a$ is a typical linear dimension of the rising bubbles. Using
parameters suitable for evolved red giants (Parker analyzed instead
the case of the Sun), K turns out to be of the order of 10$^{13} - 10^{14}$
(cgs units). Hence, large structures (100-1000 km-size) would
travel at moderate speeds, around the velocity required on the RGB, while
small instabilities (1 km-size) would provide the situation envisaged for
the AGB phases in \cite{bus07}, with high speed, close to the Alfv\'{e}n
velocity. All the cases in between these extremes are
possible. If this is the real physical situation (which fact has still
to be proven on the basis of MHD simulations), then our results would
suggest that on the RGB large magnetic domains, moving at moderate speed,
are involved in the buoyancy.

Note that, recently \citep{drake}, the association between non-convective mixing
and magnetic activity seems to have been demonstrated nicely for
the bright, very active RS CVn-type variable $\lambda$ And \citep[see for example][]{andrews}.
This star shows CNO anomalies well before reaching the
BLF. This finding, and the fact that one of our stars ($\alpha$
Boo) is known to have both a chromosphere \citep{ayr75} and
indications of photospheric magnetic fields from the Zeeman effect
\citep{senn11}, seem actually to suggest magnetic mechanisms as
very promising physical causes for driving extra-mixing in red
giants.

\begin{acknowledgements} Part of this work was supported by the
Spanish grant AYA2011-22460. S.C. acknowledges financial support
from Italian Grant FIRB 2008 "FUTURO IN RICERCA" C81J10000020001
and from PRIN-INAF 2011 "Multiple populations in Globular
Clusters: their role in the Galaxy assembly". The authors would
like to thank to K. Hinkle for providing the infrared spectra of
Aldebaran ($\alpha$ Tau). {\bf We are grateful to the refereee for very 
pertinent and useful suggestions.}
\end{acknowledgements}

\bibliographystyle{aa}
\bibliography{abiabib5}

\newpage

\begin{table*}[t!!]
\label{tabtot}
\caption{CNO abundances, C and O isotopic ratios and equivalent diffusive coefficients for the calculated
extra-mixing models.}
\centering{
\begin{tabular}{c c c c c c c c c c}
& & & & &{\large Table 4}& & & & \\
\hline  \hline
&   &       log $\epsilon$(C)   &   log $\epsilon$(N)   &   log $\epsilon$(O)   &   $^{12}$C/$^{13}$C   &   $^{16}$O/$^{17}$O   &   $^{16}$O/$^{18}$O   &   $^{17}$O/$^{18}$O   &   $D_{equiv}(cm^2s^{-1})$  \\
\hline
$\alpha$   Tau (obs) &      &   $8.25\pm$   0.12    &   8.05    &   8.48    &   10  $\pm$   2   &   1670$\pm$550    &   666$\pm$450 &   0.4 $\pm$   0.08    &   \\
            FDU     &    M$_\odot$=1.3   &   8.18    &   8.02    &   8.61    &   26.76   &   1508    &   595 &   0.39    \\
\hline
$\Delta$=   0.22    &   $\dot{M}_6$=    0.01    &   8.14    &   8.05    &   8.61    &   12.52   &   1489    &   611 &   0.41    &   1.14    $\cdot  10^{8}$ \\
$\Delta$=   0.22    &   $\dot{M}_6$=    0.012   &   8.14    &   8.04    &   8.61    &   11.96   &   1489    &   611 &   0.41    &   1.37    $\cdot  10^{8}$ \\                           $\Delta$=  0.22    &   $\dot{M}_6$=    0.015   &   8.14    &   8.04    &   8.61    &   11.53   &   1490    &   611 &   0.41    &   1.71    $\cdot  10^{8}$ \\
$\Delta$=   0.22    &   $\dot{M}_6$=    0.03    &   8.14    &   8.04    &   8.61    &   10.34   &   1490    &   611 &   0.41    &   3.42    $\cdot  10^{8}$ \\                           $\Delta$=  0.22    &   $\dot{M}_6$=    0.1 &   8.14    &   8.04    &   8.61    &   9.56    &   1490    &   612 &   0.41    &   1.15    $\cdot  10^{9}$ \\
$\Delta$=   0.22    &   $\dot{M}_6$=    0.3 &   8.14    &   8.04    &   8.61    &   9.22    &   1489    &   612 &   0.41    &   3.45    $\cdot  10^{9}$ \\                           $\Delta$=  0.2 &   $\dot{M}_6$=    0.01    &   8.1 &   8.08    &   8.61    &   11.42   &   1459    &   636 &   0.44    &   1.15    $\cdot  10^{8}$ \\
$\Delta$=   0.2 &   $\dot{M}_6$=    0.015   &   8.1 &   8.08    &   8.61    &   9.22    &   1458    &   639 &   0.44    &   1.73    $\cdot  10^{8}$ \\                           $\Delta$=  0.2 &   $\dot{M}_6$=    0.03    &   8.09    &   8.08    &   8.61    &   7.23    &   1459    &   640 &   0.44    &   3.45    $\cdot  10^{8}$ \\
$\Delta$=   0.2 &   $\dot{M}_6$=    0.1 &   8.09    &   8.07    &   8.61    &   6.07    &   1459    &   641 &   0.44    &   1.15    $\cdot  10^{9}$ \\                           $\Delta$=  0.2 &   $\dot{M}_6$=    0.3 &   8.09    &   8.07    &   8.61    &   5.76    &   1459    &   641 &   0.44    &   3.45    $\cdot  10^{9}$ \\
$\Delta$=   0.18    &   $\dot{M}_6$=    0.01    &   8.05    &   8.14    &   8.61    &   14.16   &   1383    &   694 &   0.5 &   1.15    $\cdot  10^{8}$ \\                           $\Delta$=  0.18    &   $\dot{M}_6$=    0.015   &   8.05    &   8.14    &   8.61    &   14.16   &   1383    &   694 &   0.5 &   1.73    $\cdot  10^{8}$ \\
$\Delta$=   0.18    &   $\dot{M}_6$=    0.03    &   7.99    &   8.16    &   8.61    &   5.94    &   1378    &   721 &   0.52    &   3.45    $\cdot  10^{8}$ \\                           $\Delta$=  0.18    &   $\dot{M}_6$=    0.1 &   7.97    &   8.16    &   8.61    &   4.14    &   1380    &   728 &   0.53    &   1.15    $\cdot  10^{9}$ \\
$\Delta$=   0.18    &   $\dot{M}_6$=    0.3 &   7.96    &   8.16 &
8.61    &   3.78    &   1380    &   731 &   0.53    &   3.45
$\cdot  10^{9}$ \\                           \hline $\alpha$ Boo
(obs) & &   8.06    $\pm$   0.09    &   7.67    & 8.76    &   9
$\pm$ 2 &   3030    $\pm$   530 &   1660
$\pm$   400 &   0.55    $\pm$   0.12    &   \\
    FDU & M$_\odot$=1.2   &   8.01    &   7.73    &   8.63    &   30.74   &   3341    &   1465    &   0.44    &   \\           \hline
    $\Delta$=  0.22    &   $\dot{M}_6$=    0.01    &   7.97    &   7.76    &   8.63    &   13.14   &   3248    &   1506    &   0.46    &   6.84    $\cdot  10^{7}$\\
 $\Delta$=   0.22    &   $\dot{M}_6$=    0.015   &   7.97    &   7.76    &   8.63    &   11.64   &   3248    &   1507    &   0.46    &   1.03    $\cdot  10^{8}$\\                                $\Delta$=  0.22    &   $\dot{M}_6$=    0.02    &   7.97    &   7.76    &   8.63    &   10.85   &   3235    &   1511    &   0.47    &   1.37    $\cdot  10^{8}$\\
$\Delta$=   0.22    &   $\dot{M}_6$=    0.03    &   7.97    &   7.75    &   8.63    &   10.41   &   3249    &   1507    &   0.46    &   2.05    $\cdot  10^{8}$\\                                $\Delta$=  0.22    &   $\dot{M}_6$=    0.1 &   7.97    &   7.75    &   8.63    &   9.42    &   3248    &   1508    &   0.46    &   6.84    $\cdot  10^{8}$\\
$\Delta$=   0.22    &   $\dot{M}_6$=    0.3 &   7.96    &   7.75    &   8.63    &   8.94    &   3244    &   1511    &   0.47    &   2.05    $\cdot  10^{9}$\\                                $\Delta$=  0.2 &   $\dot{M}_6$=    0.01    &   7.93    &   7.82    &   8.63    &   12.5    &   3081    &   1580    &   0.51    &   6.86    $\cdot  10^{7}$\\
$\Delta$=   0.2 &   $\dot{M}_6$=    0.015   &   7.92    &   7.81    &   8.63    &   9.46    &   3089    &   1581    &   0.51    &   1.03    $\cdot  10^{8}$\\                                $\Delta$=  0.2 &   $\dot{M}_6$=    0.03    &   7.92    &   7.81    &   8.63    &   7.28    &   3099    &   1581    &   0.51    &   2.06    $\cdot  10^{8}$\\
$\Delta$=   0.2 &   $\dot{M}_6$=    0.1 &   7.91    &   7.8 &   8.63    &   5.81    &   3093    &   1588    &   0.51    &   6.86    $\cdot  10^{9}$\\                                $\Delta$=  0.2 &   $\dot{M}_6$=    0.3 &   7.91    &   7.8 &   8.63    &   5.48    &   3094    &   1588    &   0.51    &   2.06    $\cdot  10^{9}$\\
$\Delta$=   0.18    &   $\dot{M}_6$=    0.01    &   7.88    &   7.89    &   8.63    &   17.56   &   2718    &   1736    &   0.64    &   6.89    $\cdot  10^{7}$\\                                $\Delta$=  0.18    &   $\dot{M}_6$=    0.015   &   7.84    &   7.91    &   8.63    &   11.2    &   2681    &   1788    &   0.67    &   1.03    $\cdot  10^{8}$\\
$\Delta$=   0.18    &   $\dot{M}_6$=    0.03    &   7.8 &   7.92    &   8.63    &   6.21    &   2685    &   1827    &   0.68    &   2.07    $\cdot  10^{8}$\\                                $\Delta$=  0.18    &   $\dot{M}_6$=    0.1 &   7.76    &   7.93    &   8.63    &   3.94    &   2684    &   1863    &   0.69    &   6.89    $\cdot  10^{8}$\\
$\Delta$=   0.18    &   $\dot{M}_6$=    0.3 &   7.75    &   7.94    &   8.63    &   3.54    &   2684    &   1874    &   0.7 &   2.07    $\cdot  10^{9}$\\                                \hline
    $\alpha$    Boo (obs) &     &   8.06    $\pm$   0.09    &   7.67    &   8.76    &   9   $\pm$   2   &   3030    $\pm$   530 &   1660    $\pm$   400 &   0.55    $\pm$   0.12    &   \\
FDU & M$_\odot$=1.08  &   8.03    &   7.67    &   8.63    &   32.72   &   5204    &   1393    &   0.27    &   \\
\hline  $\Delta$=   0.22    &   $\dot{M}_6$=    0.01    &   7.97    &   7.74    &   8.63    &   10.23   &   4834    &   1461    &   0.3 &   6.27    $\cdot  10^{7}$\\
$\Delta$=   0.22    &   $\dot{M}_6$=    0.015   &   7.97    &   7.74    &   8.63    &   8.73    &   4819    &   1465    &   0.3 &   9.41    $\cdot  10^{7}$\\                                $\Delta$=  0.22    &   $\dot{M}_6$=    0.03    &   7.96    &   7.74    &   8.63    &   7.53    &   4813    &   1467    &   0.3 &   1.88    $\cdot  10^{8}$\\
$\Delta$=   0.22    &   $\dot{M}_6$=    0.1 &   7.96    &   7.73    &   8.63    &   6.84    &   4817    &   1467    &   0.3 &   6.27    $\cdot  10^{8}$\\                                $\Delta$=  0.22    &   $\dot{M}_6$=    0.3 &   7.96    &   7.73    &   8.63    &   6.61    &   4815    &   1468    &   0.3 &   1.88    $\cdot  10^{9}$\\
$\Delta$=   0.2 &   $\dot{M}_6$=    0.01    &   7.9 &   7.83    &   8.63    &   9.8 &   4246    &   1582    &   0.37    &   6.29    $\cdot  10^{7}$\\                                $\Delta$=  0.2 &   $\dot{M}_6$=    0.015   &   7.89    &   7.85    &   8.63    &   7.17    &   4219    &   1599    &   0.38    &   9.44    $\cdot  10^{7}$\\
$\Delta$=   0.2 &   $\dot{M}_6$=    0.03    &   7.87    &   7.84    &   8.63    &   5.27    &   4218    &   1610    &   0.38    &   1.89    $\cdot  10^{8}$\\                                $\Delta$=  0.2 &   $\dot{M}_6$=    0.1 &   7.86    &   7.84    &   8.63    &   4.27    &   4221    &   1617    &   0.38    &   6.29    $\cdot  10^{8}$\\
$\Delta$=   0.2 &   $\dot{M}_6$=    0.3 &   7.85    &   7.84    &   8.63    &   4.03    &   4213    &   1622    &   0.38    &   1.89    $\cdot  10^{9}$\\                                $\Delta$=  0.18    &   $\dot{M}_6$=    0.01    &   7.83    &   7.93    &   8.63    &   14.33   &   3103    &   1859    &   0.6 &   6.31    $\cdot  10^{7}$\\
$\Delta$=   0.18    &   $\dot{M}_6$=    0.015   &   7.78    &   7.95    &   8.63    &   8.34    &   3232    &   1902    &   0.59    &   9.47    $\cdot  10^{7}$\\                                $\Delta$=  0.18    &   $\dot{M}_6$=    0.03    &   7.69    &   7.99    &   8.63    &   4.88    &   2986    &   2102    &   0.7 &   1.89    $\cdot  10^{8}$\\
$\Delta$=   0.18    &   $\dot{M}_6$=    0.1 &   7.59    &   8.03    &   8.63    &   3.27    &   2917    &   2264    &   0.78    &   6.31    $\cdot  10^{8}$\\                                $\Delta$=  0.18    &   $\dot{M}_6$=    0.3 &   7.57    &   8.04    &   8.63    &   3.04    &   2945    &   2283    &   0.78    &   1.89    $\cdot  10^{9}$\\
\hline
\end{tabular}
}
\end{table*}

\end{document}